\documentclass[preprint,amsmath,amssymb,aps,prb]{revtex4-1}
\usepackage{graphicx}
\usepackage[colorlinks=true, allcolors=blue]{hyperref}

\begin{document}
\title{A new strategy for directly calculating the minimum eigenvector of matrices without diagonalization}
\author{Wei Pan, Jing Wang and De-Yan Sun}
\affiliation{Department of Physics, East China Normal University, 200241 Shanghai, China}

\begin{abstract}
The diagonalization of matrices may be the top priority in the application of modern physics. In this paper, we numerically demonstrate that, for real symmetric random matrices with non-positive off-diagonal elements, a universal scaling between the eigenvector and matrix elements exists. Namely, each element of the eigenvector of ground states linearly correlates with the sum of matrix elements in the corresponding row. Although the conclusion is obtained based on the random matrices, the linear relationship still keeps for regular matrices, in which off-diagonal elements are non-positive. The relationship implies a straightforward method to directly calculate the eigenvector of ground states for a kind of matrices. The test on both Hubbard and Ising models shows that, this new method works excellently.  
\end{abstract}

\maketitle
	
Without any doubt, a lot of scientific problems are directly related to matrix algebra. Obtaining eigenvalues or eigenvectors of matrices is one of the basic tasks in many fields of science and technology. The significance of matrices is especially prominent in modern physics. Almost all quantum problems come down to the diagonalization of matrices in principle. A Hamiltonian matrix contains all the information of a corresponding quantum system, and the density matrix reflects all thermal properties of a system at finite temperatures. However the diagonalization of an arbitrary matrix for many physical systems of practical interests is definitely not an easy stuff. Especially in condensed matter and statistical physics, the number of particles usually has the magnitude of the Avogadro constant, correspondingly the many-body Hamiltonian matrix may quickly become a hopeless scale, which is too large to be diagonalized by any conventional mathematical methods.

Over half a century, great efforts have been carried on the matrix diagonalization, and remarkable progresses have been made associated with the rapid development of modern computational technology. Speaking limited to physics, many diagonalization methods have been proposed, such as, exact diagonalization method \cite{Lanczos1950,PhysRevLett.72.1545,Zhang2010,Si1994PRL,RevModPhys.68.13}, quantum Monte-Carlo \cite{Foulkes2001RMP,vonderLinden1992,Kolorenc2011,RevModPhys.83.349}, and the density matrix renormalization group \cite{White1992PRL,Schollwuck2005RMP,RevModPhys.80.395,Schollwuck2011AP} $etc$.  Even so, for almost all real physics systems, the direct diagonalization of matrices is still an impossible mission even with the help of modern computers.

 When the direct diagonalization becomes impractical tasks, one naturally ask whether other feasible methods exist? Obviously, for a diagonalizable matrix, both eigenvalues and eigenvectors are uniquely determined by its elements. One possible idea is to establish an immediate connection between eigenvectors and matrix elements. If this kind of connection can be figured out, it may be an appealing method for matrix diagonalization. Certainly, it is not easy to obtain the possible connection, because few precedents can be followed to realize this idea. Fortunately, the strategy adopted in Big Data analysis and Machine Learning is heuristic, which can be used for reference. In Big Data analysis and Machine Learning, it usually makes predictions or decisions based on vast data sets without being programmed to perform the task. For examples, in recent studies on many-body quantum systems, the physical properties are predicted without explicitly diagonalizing Hamiltonian matrices \cite{Cai2018PhysRevB,Chng2017PRX}.

Enlightened by many successful cases in the machine learning or Big Data analysis, we expect that, the connection between matrix elements and eigenvectors could be pried through the deep analysis for an enormous number of matrices. Here as a first attempt, we have focused on the random matrices (RMs), which are introduced in 1955 by Wigner \cite{Wigner1955}. RMs are common and important in many fields of physics. In quantum chaos, the Bohigas–Giannoni–Schmit conjecture is closely related to RMs \cite{Bohigas1984PRL}. In quantum optics, transformations described by RMs are crucial for demonstrating the advantage of quantum over classical computation (see, e.g., in Ref. \cite{Aaronson2013,Russell2017}). In condensed matter physics,  the fractional quantum Hall effect \cite{Callaway1991PRB},  Anderson localization \cite{Janssen2000PRE}, quantum dots \cite{Zumbuhl2002PRL},  superconductors \cite{Bahcall1996PRL}, spin glasses \cite{AndersonSPGS1,AndersonSPGS2} are connected to RMs too. 
And more multi-applications of RMs in physics can be found, for instance, in Refs. \cite{Guhr1998,Deutsch2018Review,PhysRevE.50.888,RevModPhys.53.385,Gomez2011Review,Shen2008PRC}.

In this work, we have systematically studied random real symmetric matrices with non-positive off-diagonal elements (hereafter labeled as RRSMs). If the value of matrix element is interpreted as the scattering amplitude as what does in quantum physics, RRSMs describe a system having the random scattering amplitude. 
The choice of random matrices is also in order to make our conclusion having the universality, since random matrices cover all the possibility in principle.
At least for this kind of matrices, we have found a strong correlation between the eigenvector of the minimum eigenvalue (EME) and the sum of matrix elements (SME) in corresponding row. This result implies a new method for diagonalizations of RRSMs regardless matrix dimensions. The achievement of this work can also shed light on the solution of other complex matrices.

A RRSM is denoted by $H$, and its dimension by $N$. $H_{ij}$ represents the matrix element in the $i$-th row and the $j$-th column. For the sake of convenient, we assume that $H$ is a Hamiltonian matrix of a certain quantum system. This assumption is just for convenience, does not alter our conclusion. Assuming orthogonal complete basis being $|e_i>$, where the subscript $i$ refers the $i$-th basis,the projection of $|e_i>$ on EME ($|G>$ ) reads as $g_i$, $i.e$, $g_i$=$<G|e_i>$ and $|G>=\sum_ig_i|e_i>$. Thus, the minimum eigenvalue ($<G|H|G>$) is $\sum_{i,j}H_{ij}g_ig_j$. The formula of $<G|H|G>$ hints that the SME in each row ($S_i=\sum_jH_{ij}$) should reflect the contribution of $|e_i>$ in the ground state. We intuitively conjecture that the coefficients ($g_i$) are correlated to $S_i$, and the smaller ($i.e.$, the more negative) value of $S_i$ corresponds to the larger value of $g_i$. In the following context, we will demonstrate this conjecture and find the generic relationship. For better comparison among all matrices, $g_i$ is re-scaled according to the normalization condition, and $S_i$ is also normalized based on $\sum_iS_i^2=1$.

The dimension of RRSMs considered in this work ranges from 100 to 10000. Around ten thousand matrices are calculated for each dimension. All matrix elements are generated according to two types of distribution, $i.e,$ the uniform and Gaussian distribution. For the uniform distribution, a number in the range of [$X_{min}$,0], where $X_{min}$ is a negative number, is randomly chosen and assigned to a matrix element. For the Gaussian distribution, a matrix element is generated through the Gaussian distributions, in which the Gaussian variation and mean value are set as 1.0 and $-2.0$ respectively. In Gaussian distribution, a few percent of off-diagonal elements are positive, however this situation does not alter our conclusions. To further check the universality of our results, the matrix elements in several rows are also randomly enlarged or reduced. And the effect of the matrix density ($\rho$) (e.g., the number of non-zero elements divided by the total number of elements) is also investigated. To reduce the matrix density, a certain randomly chosen matrix elements are taken to be zero. For all matrices, the ground states are obtained by direct diagonalization first, then $g_i$ and $S_i$ are calculated straightforwardly. 

\begin{figure}
\includegraphics[width=0.75\textwidth]{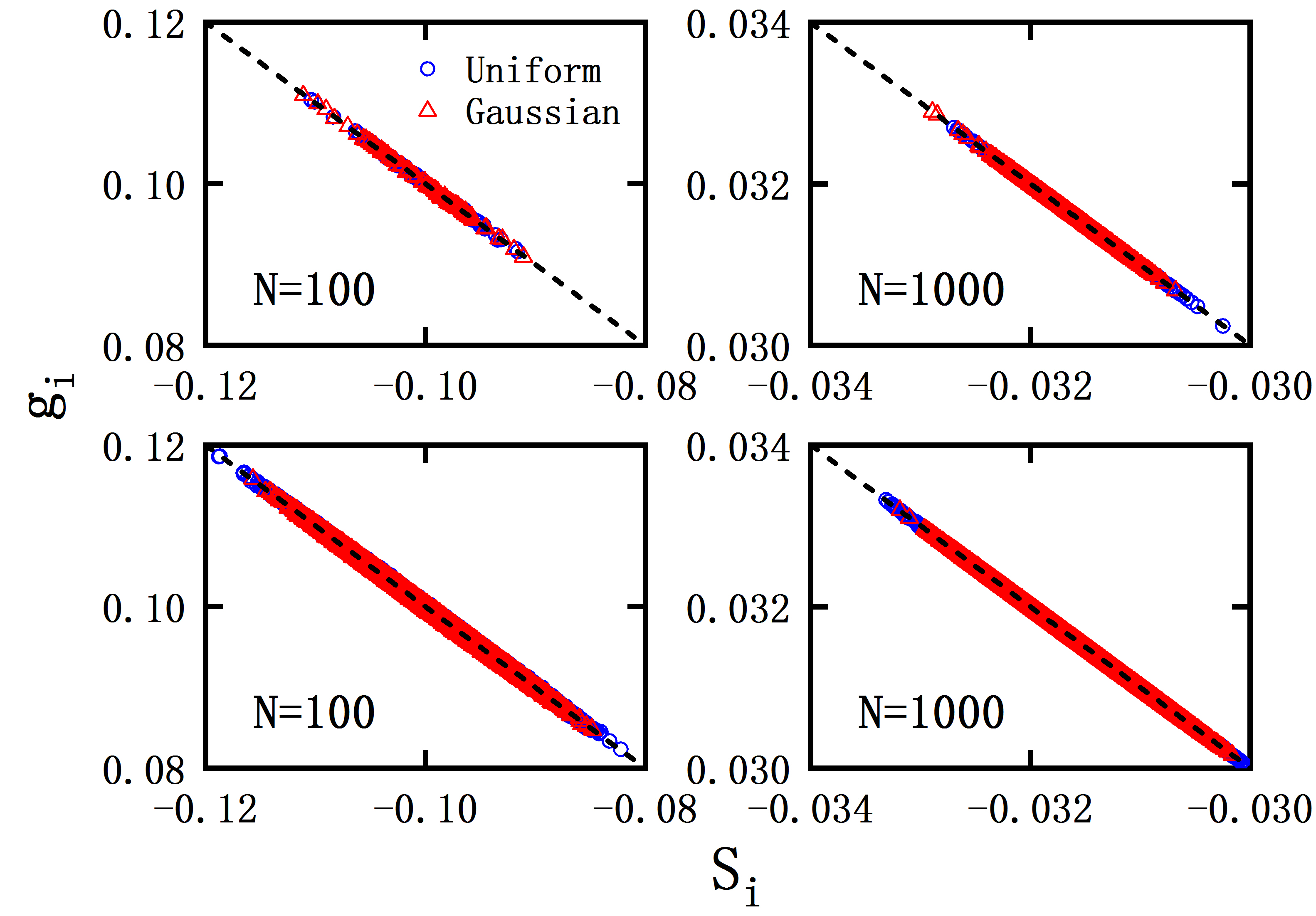}
\caption{\label{fig:Figure1} (Color online) Elements of eigenvector for the minimum eigenvalue ($g_i$) versus the sum of matrix elements in corresponding row ($S_i$). Upper panel: one arbitrarily chosen matrix with uniform (circle) and Gaussian distribution (triangle). Lower panel: all studied matrices with uniform (circle) and Gaussian distribution (triangle). For all cases, an evident linear relationship can be observed.}
\end{figure}

Fig.~\ref{fig:Figure1} shows the element of EME ($g_i$) versus SME in corresponding row ($S_i$) for both uniform (circle) and Gaussian distribution (triangle), where all matrices are dense ($\rho\sim 1$). In the upper panel of this figure, the results for an arbitrarily chosen RRSM with the dimension of 100 (left) and 1000 (right) are presented. It can be seen that, $g_i$ increases with the increase of $S_i$. The correlation between $g_i$ and $S_i$ shows an almost perfect linearity. The solid lines are the best fitting to the data, which produce a slope of $-1$ and intercept of $0$. This correlation does not merely appear in several individual matrices, but in all the studied matrices, as shown in the lower panel of Fig.~\ref{fig:Figure1}. It needs to point out that, although Fig.~\ref{fig:Figure1} only presents the results of RRSMs with two specific dimensions (N=100 and 1000), all the studied RRSMs exhibit the same linear scaling.

\begin{figure}
\centering
\includegraphics[width=0.75\textwidth]{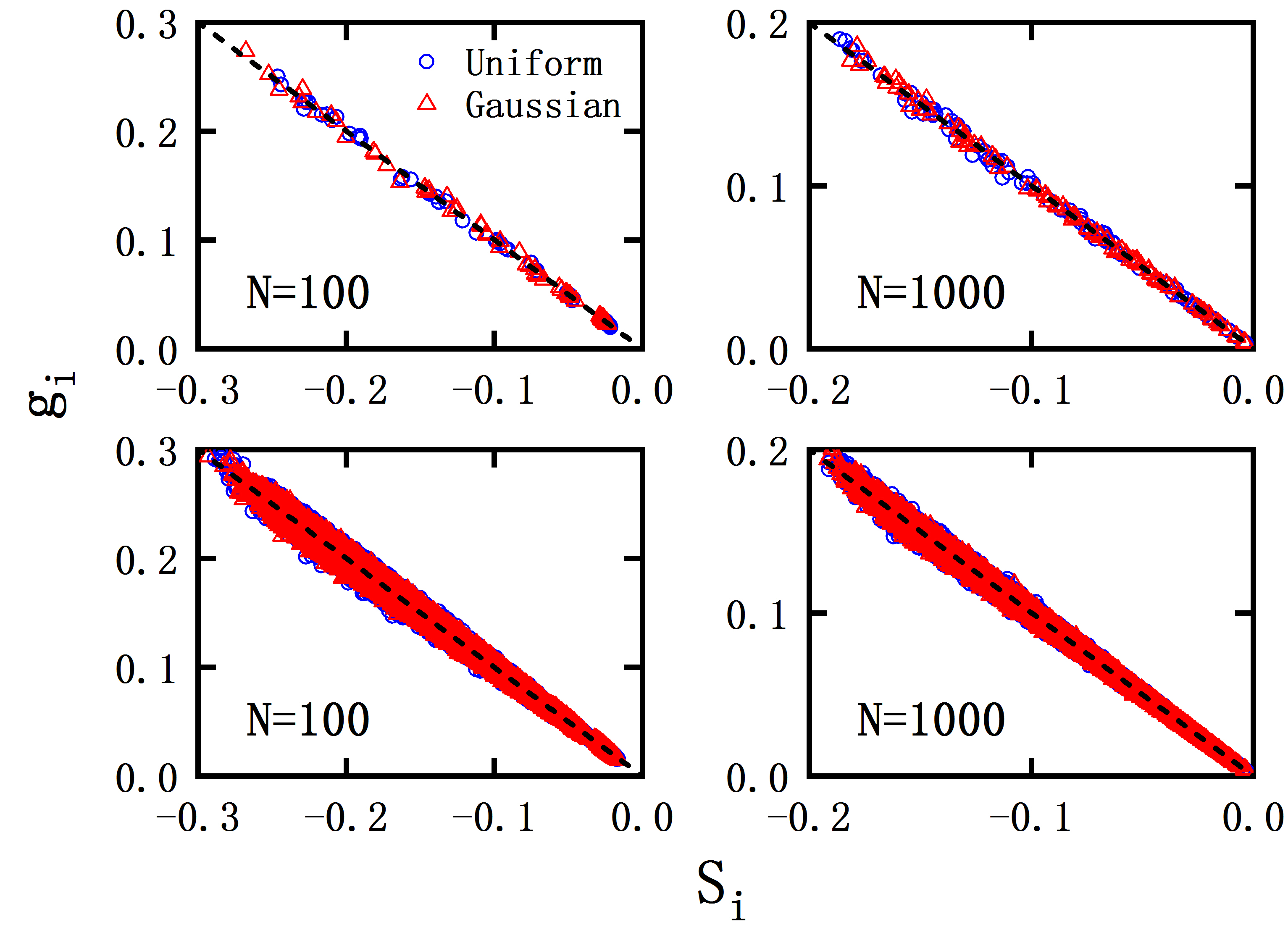}
\caption{\label{fig:Figure2} (Color online) Similar to Fig.~\ref{fig:Figure1}, but the amplitude of matrix elements in some rows is enlarged or reduced. In this case, the linear relationship is still kept.}
\end{figure}

It is easy to demonstrate that, this linear relationship will not break down if the magnitude of all matrix elements changes simultaneously. The next question is, what would happen if the magnitude of matrix elements in some rows is significantly larger or smaller than those in other rows? Our results show that this linear relationship still keeps. In Fig.~\ref{fig:Figure2}, the results are extended to this kind of special cases, in which matrix elements in several rows are much larger or smaller than others. These points corresponding to the enlarged (reduced) rows locate in the top (bottom) area in Fig.~\ref{fig:Figure2}. One can see that, the correlation still remains linearity with a slope of $-1$ and intercept of $0$. And the linear behavior does not break down no matter how many rows being enlarged or reduced.

From Fig.~\ref{fig:Figure1} and Fig.~\ref{fig:Figure2}, we notice a remarkable feature, namely the linear relationship does not show evident deviation when the dimension of matrices increases. This point is extremely important for practical applications, because we are not only interested in the linear relationship, but also concern the possible application of this relationship for high dimensional matrices. Obviously we can not extend our calculations to all high dimensional matrices. However, if we know how the deviation from linearity changes with the increase of matrix dimensions, we are able to predict the validity at higher dimensions. To check this point, the root-mean-squared ($rms$) deviation  from linearity is calculated, which reads:   
\begin{equation}
rms=\sqrt{\frac{\sum_i^N\left(g_i-(-S_i)\right)^2}{N}}
\end{equation}
where $N$ is the dimension of matrices. 

\begin{figure}
\centering
\includegraphics[width=0.75\textwidth]{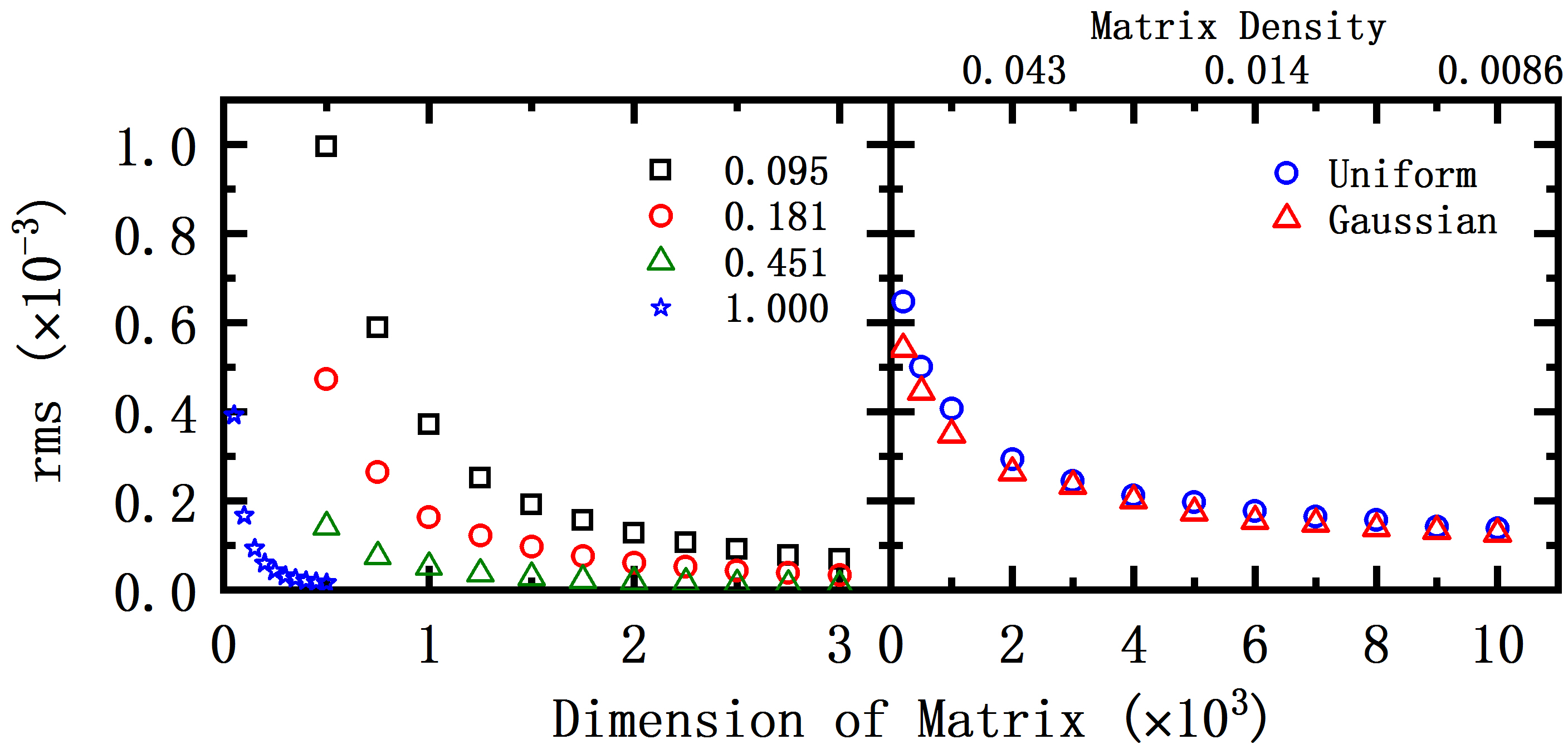}
\caption{\label{Figure3} (Color online) The root-mean-squared deviation from linearity ($rms$) as the function of matrix dimension. Left panel: for matrices with fixed density ($\rho$) of $1.000$ (five-pointed star), $0.095$ (square), $0.181$ (circle) and $0.451$ (triangle). Right panel: matrices with density ($\rho$) varying as $\sim\frac{1}{N}$, where $N$ is the matrix dimension. For all cases, the deviation decreases dramatically as the increase of matrix dimension, and quickly stabilizes at a small value.}
\end{figure}

In the left panel of Fig.~\ref{Figure3} we have depicted $rms$ deviations from linearity as function of the matrix dimension both for dense and sparse matrices, where the density of matrices is chosen as $1.000$ (five-pointed star), $0.095$ (square) $0.181$ (circle) and $0.451$ (triangle). It is encouraging that, both dense and sparse matrices exhibit the similar trend, namely $rms$ deviation decreases dramatically as the dimension of matrices increasing, and quickly stabilizes at a very small value. At the same dimension (N), $rms$ deviation from linearity for sparse matrices is a little bit larger than that for dense ones. 

In many physical systems of practical interests, the density of matrix decays with its dimension as $\rho\sim\frac{1}{N}$. The right panel of Fig.~\ref{Figure3} shows the result for this kind of matrices. It is more encouraging that the deviation is still convergent at large matrix dimensions in this case. Although we can not examine all the matrices with non-positive elements, the matrices considered in this work are quite general and sufficiently large in amounts, we believe that the linear relationship should be universal for this kind of matrices. We can conjecture in advance that the linear relationship may become strict as the matrix dimension approaching to infinite.

It needs to be pointed out that, although diagonal elements are generated using the same way as off-diagonal ones, the above conclusions are irrelevant to the sign of diagonal elements. It can be proved as follows: Suppose $\bar{H}$ being a random real symmetric matrix with non-positive off-diagonal elements but arbitrary diagonal elements, $\bar{H}$ can always be expressed as $H_r+D\times I$. Here $H_r$ is a RRSM discussed above, $I$ is a unity matrix, and $D$ is a arbitrary constant. It is self-evident that, $\bar{H}$ and $H_r$ share the same eigenvectors, since $D\times I$ just simultaneously shifts all diagonal elements. Thus the linear relationship discussed above holds for $\bar{H}$ if $H_r$ does.

The linear relationship presented in RRSMs should have both physical and mathematical origin. Unfortunately, neither simple mathematics theorem nor physical theory can provide definite information regarding this issue. However, we still can make some arguments from both physics and mathematics considerations as follows.

For RRSMs, we are able to easily prove that all $g_i$  have the same sign by using the Perron-Frobenius theorem \cite{Perron1907,Frobenius1908}. For convention, we take $g_i\geq 0$. Under the mean-field-like approximation, the energy of ground state $\sum_{i,j}H_{ij}g_ig_j$ can be written as $\sum_{i,j}H_{ij}g_i<g>$, where $<g>$ refers the mean value. Since the ground state has the minimum energy, $g_i$ should be proportional to $\sum_{i,j}H_{ij}$, $i.e.$, $S_i$.

As the further remark, we should discuss in which conditions the linear relationship will be broken down. First, for diagonal dominated matrices, in which the distribution width of diagonal elements is larger than the sum of off-diagonal ones, the linear relationship breaks down, but the positive correlation keeps. Second, for some band matrices with negative off-diagonal elements, again the linear relationship breaks down, but the positive correlation holds. It needs to be addressed that, although the current results are obtained for RRSMs, our preliminary results show that, the positive correlation between $g_i$ and $S_i$ may still be kept if  the sum of negative elements prevails over the sum of positive ones in a matrix. Surely the correlation may be complicated rather than simple linearity. For more general cases, the further investigations are worth doing \cite{Pan2}.

The linear relationship obtained in present work should have broad applications for many physics systems of practical interest, in which the Hamiltonian matrices are similar to RRSMs. The applications are manifold. It can be used to determine the energy and wave function of ground states, or to analyze the physical properties of ground states. This strategy has the advantage of briefness, high efficiency and is easily generalized to arbitrary large dimensions. Of course, the most powerful application maybe combines the linear relationship with other modern matrix diagonalization techniques, such as Monte Carlo method, to calculate the properties of ground state. 

Although the conclusion is obtained based on the random matrices, the linear relationship still keeps for regular matrices, in which off-diagonal elements are non-positive. As practical examples, we have tested the linear relationship on both one-dimensional Hubbard model\cite{Hubbard1963} and quantum Ising model \cite{Sachdev_book}. The Hubbard model is extremely fundamental and important for a variety of areas, especially in the study of strongly correlated quantum systems. It is an important model to describe metal-insulator transitions\cite{Imada1998} and to understand high temperature superconductors\cite{Zhang1988,Orenstein2000}. The quantum Ising model, or equivalently, the Ising model in a transverse field, is one of the most widely used paradigm in studying quantum phase transitions.  

For the one-dimensional 4-site Hubbard model, the Hamiltonian matrix is a $36\times36$ one, which can be directly diagonalized. If anti-periodic boundary conditions are used, all off-diagonal elements are non-positive. The on-site coupling strength is chosen to be $U/t=0$ and $U/t=1$ respectively. For the larger $U$ situations, it corresponds to diagonally dominant matrices, which is beyond the scope of current work. For the one-dimensional quantum Ising model,\cite{Sachdev_book} the Hamiltonian reads	$H_I=-g\sum_i\hat{\sigma}_i^x-\sum_{<ij>}\hat{\sigma}_i^z\hat{\sigma}_j^z$, 
where $\hat{\sigma}_i^x$ and $\hat{\sigma}_i^z$ are Pauli matrices and $g>0$ is a dimensionless parameter. In the basis where $\hat{\sigma}_i^z$ is diagonal, the off-diagonal elements of the Hamiltonian matrix of $H_I$ are constituted by $-g$ and 0. The large $g$ corresponds to the case, in which the Hamiltonian matrix is not diagonally dominant and the linear relationship should be kept. In current studies, $g=10$ and periodic boundary conditions are adopted. The system size (length of chain) considered includes $L=4,6,8,10,12,14$. In order to use the linear relationship, we assume $g_i=c_1S_i+c_2=c_1(S_i+c_2/c_1)$, obviously only $c_2/c_1$ is important for physical wave functions, thus the energy of ground state can be obtained by means of variation respect to the only parameter $c=c_2/c_1$.

\begin{figure}
\centering
\includegraphics[width=0.6\textwidth]{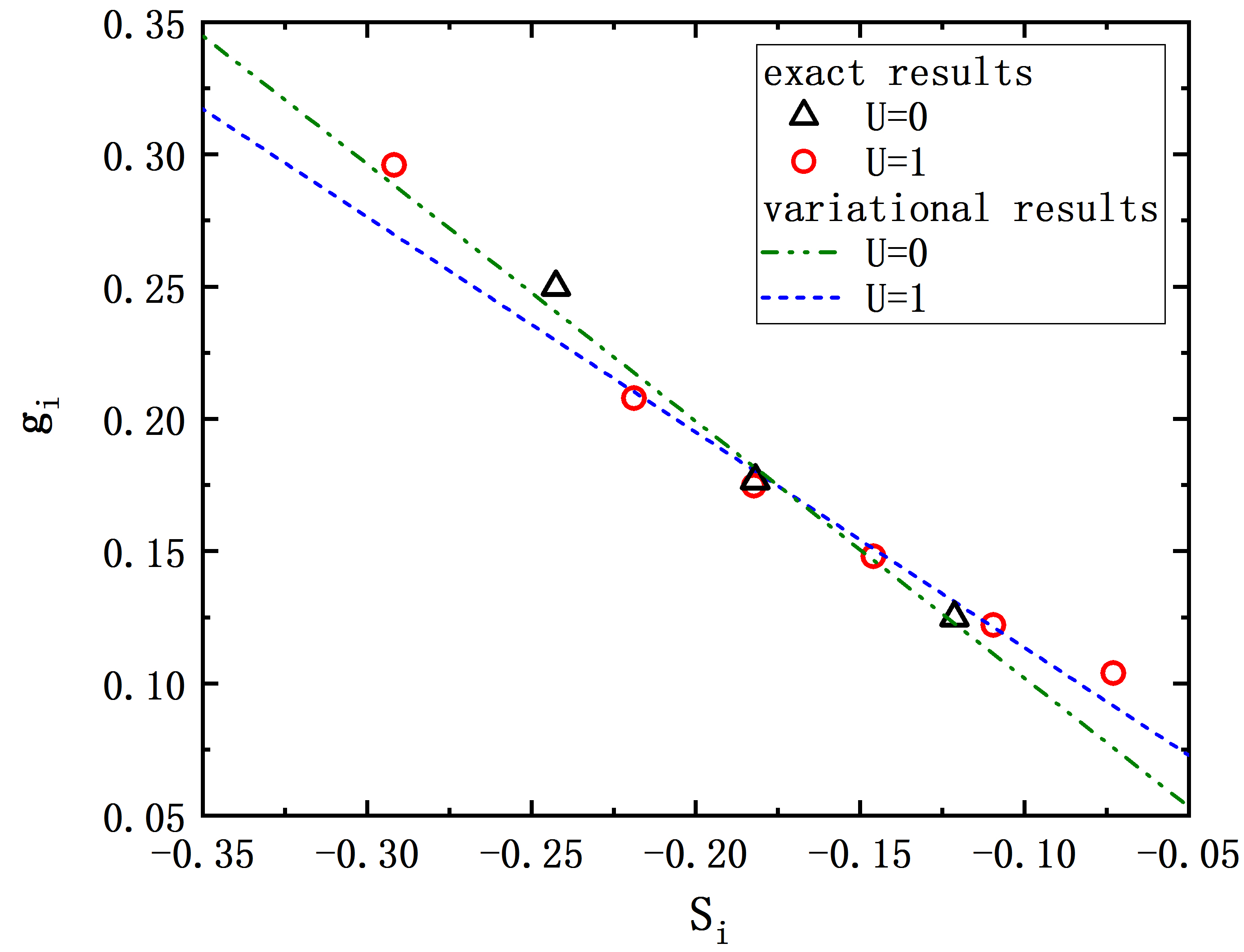}
\caption{\label{Figure4} (Color online) Element of the eigenvector for the minimum eigenvalue ($g_i$) versus the sum of matrix elements in corresponding row ($S_i$) for one-dimensional 4-site half-filled Hubbard model. The triangles and circles correspond to the coupling strength of $U/t=0$ and $U/t=1$ respectively. The dash-dot-dot and dashed lines are variational results based on linear relationship for $U/t=0$ and $U/t=1$ respectively. Approximate linear relationships can be observed, which are in agreement with our predication.}
\end{figure}

Fig.~\ref{Figure4} presents the coefficients of wave-function of ground state for the Hubbard model, in which the dashed lines and symbols are the variational and exact values respectively. One can see that, although the corresponding matrices for both $U/t=0$ and $U/t=1$ cases are not random, the clear linear relationship between $g_i$ and $S_i$ holds, which is in agreement with our predication. And the ground-state energy obtained according to the linear relationship is much close to the exact value as shown in Table \ref{table1}.

\begin{figure}
\centering
\includegraphics[width=0.8\textwidth]{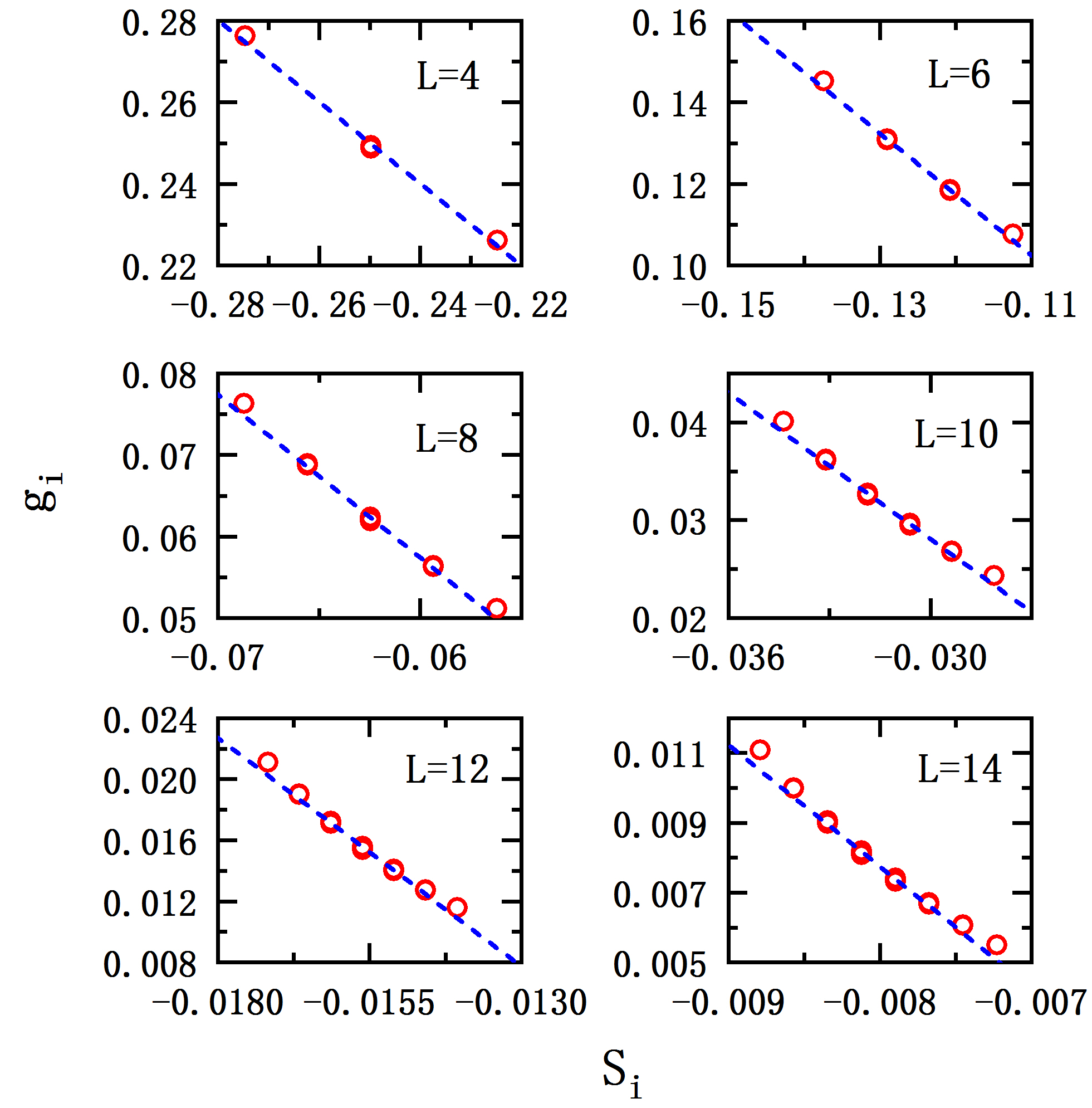}
\caption{\label{Figure5} (Color online) Similar to Fig. \ref{Figure4}, but the results are corresponding to the one-dimensional quantum Ising model for system sizes of $L=4,6,8,10,12,14$ respectively. The dashed lines are variational results based on linear relationship, and the symbols represent the exact results.}
\end{figure}

Similar to Fig.~\ref{Figure4}, Fig.~\ref{Figure5} presents the results for the one-dimensional quantum Ising model. Clearly, the linear correlation between $g_i$ and $S_i$ is still pronounced. It can be seen that, the ground state energy obtained according to the linear relationship is quite accurate as listed in Table~\ref{table1}, the errorbar is less than 0.01\%.
  
\begin{table}
\caption{\label{table1}The results of variational parameter ($c$) and calculated ground state energy for both the one-dimensional Hubbard and the quantum Ising model. The ground state energies obtained according to the linear relationship ($E_{\rm{scaling}}$) are much close to the exact value ($E_{\rm{exact}}$). Here the exact energy is obtained by direct diagonalization.}
\centering
\begin{tabular}{|c|c|c|c|c|c|c|c|c|}
\hline
 & \multicolumn{6}{|c|}{Quantum Ising Model} & \multicolumn{2}{|c|}{Hubbard Model}\\
\hline
       & $L=4$ & $L=6$   & $L=8$   & $L=10$  & $L=12$  & $L=14$  & $U=0$ & $U=1$\\
\hline 
$c$   & 0.000620 & -0.0413 & -0.0311 & -0.0187 & -0.0104 & -0.00556 & 0.00954 & -0.0137\\
\hline 
$E_{\rm{scaling}}$ & -10.024938 & -10.024907 & -10.024876 & -10.024845 & -10.024815 & -10.024785 & -1.41202 & -1.17314\\ 
$E_{\rm{exact}}$ & -10.024938 & -10.025015 & -10.025016 & -10.025016 & -10.025016 & -10.025016 & -1.41421 & -1.18082\\ 
\hline 
\end{tabular}
\end{table}

In summary, we have explored the probability to establish an immediate connection between the eigenvector and matrix elements. At least, for real symmetric random matrices with non-positive off-diagonal elements, this kind of connection has been figured out. Namely, the eigenvector of ground states can be directly obtained by the sum of matrix elements in each row. This connection provides a feasible method to calculate the eigenvector of matrices without diagonalization. Although the linear relationship is obtained based on the random matrices, the test on model systems further confirms the validity of the scaling between the ground state eigenvector and matrix elements.

Acknowledgments
Project supported by the National Natural Science Foundation of China (Grant No. 11874148). The computations were supported by ECNU Public Platform for Innovation (001).

\bibliography{ref}
\clearpage

\end{document}